\documentclass[conference]{IEEEtran}
\IEEEoverridecommandlockouts
\usepackage{cite}
\usepackage{amsmath,amssymb,amsfonts}
\usepackage{algorithmic}
\usepackage{algorithm}
\usepackage{graphicx}
\usepackage{textcomp}
\usepackage{lipsum}
\usepackage{xcolor}
\usepackage{booktabs,multirow,bm}
\usepackage{tabularx}
\def\BibTeX{{\rm B\kern-.05em{\sc i\kern-.025em b}\kern-.08em
    T\kern-.1667em\lower.7ex\hbox{E}\kern-.125emX}}

\usepackage{graphicx}
\ifCLASSOPTIONcompsoc
    \usepackage[caption=false, font=normalsize, labelfont=sf, textfont=sf]{subfig}
\else
\usepackage[caption=false, font=footnotesize]{subfig}
\fi

\newcommand{\x}{\mathbf{x}}
\newcommand{\z}{\mathbf{z}}
\newcommand{\bc}{\mathbf{c}}
\newcommand{\y}{\mathbf{y}}

\newcommand{\f}{\mathbf{f}}

\newcommand{\W}{\mathbf{W}}
\newcommand{\bb}{\mathbf{b}}
\newcommand{\hz}{\hat{\mathbf{z}}}
\newcommand{\hy}{\hat{\mathbf{y}}}
\newcommand{\hZ}{\hat{Z}}

\newcommand{\bR}{\mathbb{R}}

\newcommand{\bPhi}{\boldsymbol{\phi}}
\newcommand{\bNu}{\boldsymbol{\nu}}
\newcommand{\bTheta}{\boldsymbol{\theta}}

\newcommand{\bPsi}{\boldsymbol{\psi}}

\DeclareMathOperator*{\argmin}{argmin}

\begin{document}

\title{ {Deep Learning-Based Adaptive Joint Source-Channel Coding using Hypernetworks}
}

\author{\IEEEauthorblockN{Songjie~Xie\IEEEauthorrefmark{1}, Hengtao~He\IEEEauthorrefmark{1}, Hongru~Li\IEEEauthorrefmark{1}, Shenghui~Song\IEEEauthorrefmark{1},~Jun~Zhang\IEEEauthorrefmark{1},~\IEEEmembership{Fellow,~IEEE},} 
\IEEEauthorblockN{Ying-Jun~Angela~Zhang\IEEEauthorrefmark{2},~\IEEEmembership{Fellow,~IEEE}, and Khaled~B.~Letaief\IEEEauthorrefmark{1},~\IEEEmembership{Fellow,~IEEE}}
\IEEEauthorblockA{\IEEEauthorrefmark{1}Dept. of Electronic and Computer Engineering, The Hong Kong University of Science and Technology, Hong Kong
}
\IEEEauthorblockA{\IEEEauthorrefmark{2}Dept. of Information Engineering, The Chinese University of Hong Kong, Hong Kong}
Email: \{sxieat, hlidm\}@connect.ust.hk, \{eehthe, eeshsong, eejzhang, eekhaled\}@ust.hk, yjzhang@ie.cuhk.edu.hk
}

\maketitle

\begin{abstract}
Deep learning-based joint source-channel coding (DJSCC) is expected to be a key technique for  {the} next-generation wireless networks. However, the existing DJSCC schemes still face the challenge of channel adaptability as they are typically trained under specific channel conditions. In this paper, we propose a generic framework for channel-adaptive DJSCC by utilizing hypernetworks. To tailor the hypernetwork-based framework for communication systems, we propose a memory-efficient hypernetwork parameterization and then develop a channel-adaptive DJSCC network, named Hyper-AJSCC. Compared with existing adaptive DJSCC based on the attention mechanism, Hyper-AJSCC introduces much fewer parameters and can be seamlessly combined with various existing DJSCC networks without any substantial modifications to their neural network architecture. Extensive experiments demonstrate the better adaptability to channel conditions and higher memory efficiency of Hyper-AJSCC compared with state-of-the-art baselines.
\end{abstract}


\section{Introduction}
With the widespread success and effectiveness of artificial intelligence (AI) in various domains, it is anticipated that  {the} next-generation communications will be revolutionized by an in-depth integration of advancements in AI technology~\cite{letaief2019roadmap, 9018199, 10143629, 10278093}.  
Recently, deep learning-based joint source-channel coding (DJSCC) schemes have been developed to support ubiquitous connected intelligent devices and AI services. 
Compared with the conventional separate source-channel coding strategy, the recently proposed DJSCC schemes exhibit numerous appealing properties, including substantial enhancements in data reconstruction~\cite{bourtsoulatze2019deep} and graceful performance loss with degrading channel quality~\cite{kurka2020deepjscc}. 
These advantages not only benefited conventional data-oriented communication systems but also facilitated emerging paradigms such as semantic and task-oriented communication~\cite{xie2021deep, dai2022nonlinear, shao2021learning, 10159007, 10183789, li2023task}.

Despite the empirical success of the existing DJSCC schemes, they still face the challenge of channel adaptability. In particular, the existing DJSCC schemes proposed for image and text transmission typically adopt a deep learning-based autoencoder with a non-trainable layer to simulate the noisy channel. Then, the encoder and decoder are optimized in an end-to-end manner under particular channel conditions. Consequently, the mismatch between the channel conditions during the training and deployment stages results in a significant performance degradation~\cite{bourtsoulatze2019deep}. One solution is to train multiple networks for different channel conditions and select the suited one for data transmission according to the channel conditions. However, this strategy requires {an} extremely high storage memory to store multiple networks, thereby limiting its practical applications, especially in resource-constrained scenarios. The ideal approach for solving the adaptability challenge of DJSCC is to design a single network that can adjust its parameters to different channel conditions. To achieve this goal, recent research~\cite{xu2021wireless} proposed to employ attention mechanisms to adjust intermediate features according to channel conditions, and has been extended to various DJSCC scenarios~\cite{yang2022deep, 9878262}. More recently, the transformer-based method~\cite{wu2022vision} was utilized to leverage the intrinsic self-attention mechanism to improve channel adaptability. 
However, the attention-based approaches suffer from several limitations when integrated into existing DJSCC schemes.
Firstly, attention mechanisms can be computationally expensive due to their involvement in the computation of multiple neural network layers. 
On the other hand, introducing attention modules to existing DJSCC networks disrupts their original structures. This disruption impacts the overall design and functionality of the original DJSCC networks, which can result in compatibility issues.

To address these problems, we propose a generic channel adaptive framework that can be directly employed in any DJSCC network to enhance their adaptability to channel variations.
Hypernetworks~\cite{ha2016hypernetworks, bae2022multi} have been proven effective in enhancing the flexibility and adaptability of neural networks due to their ability to generate network parameters dynamically.
Thus, we adopt them to generate the parameters of {the} optimal encoder/decoder according to the channel conditions.  
 {To make the proposed hypernetwork-based framework compatible with communication systems,}
we propose a scalable and memory-efficient hypernetwork parameterization and further develop a channel-adaptive DJSCC scheme, {namely} Hyper-AJSCC.
 {The proposed Hyper-AJSCC approach offers several advantages over attention-based methods. It introduces significantly fewer parameters and can be seamlessly integrated into a wide range of existing DJSCC schemes. More importantly, this integration does not compromise the consistency or fundamental structure of the original neural network backbones.}
To verify the performance of our method in data/task-oriented communication paradigms, we conduct extensive experiments on image transmission and classification tasks. The results demonstrate the advantages of the proposed Hyper-AJSCC schemes in improving the adaptability of the network to channel conditions.

\section{System Model and Problem Formulation}
	\label{sec:model_des}
  \begin{figure}[t]
		\centering
		\includegraphics[width=\linewidth]{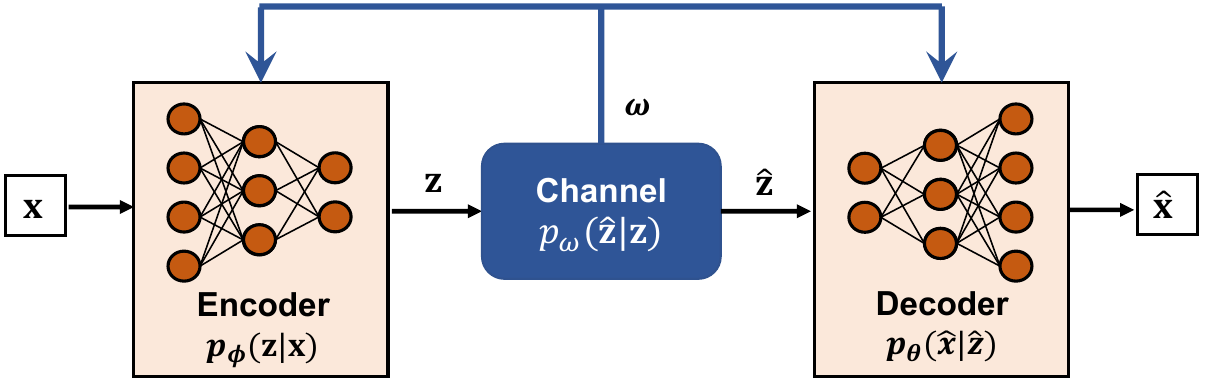}
		\caption{The considered system model of point-to-point communication with known channel condition at the transmitter and receiver.
		}
		\label{fig:system-model_Hyper}
	\end{figure}
	\subsection{System Model}
	\label{subsec:system_model}
 As shown in Fig.~\ref{fig:system-model_Hyper}, we consider a point-to-point DJSCC system consisting of a single encoder and a decoder, both of which are parameterized by neural networks. The channel conditions are known by both the DJSCC encoder and decoder. It constitutes a general model that can be implemented in data-oriented and task-oriented communication systems. A data source generates the observable data $\x \in \bR^n$ with distribution $p(\x)$ and forms a dataset $\{\x^{(i)}\}_{i=1}^N$ consisting of $N$ independent and identically distributed (i.i.d.) data points.
At the transmitter side, the input data $\x$ is encoded into $d$-dimensional channel input symbols $\z \in \mathbb{C}^d$ using a probabilistic encoder $p_{\bPhi}(\z|\x)$ parameterized by the trainable parameters $\bPhi$ of neural networks.
	
	{The} encoded representation $\z$ is {then} transmitted over a noisy channel. In this work, we consider the additive white Gaussian noise (AWGN) channel. The additive noise vector $\boldsymbol{\epsilon}$ is sampled from an isotropic Gaussian distribution with variance $\sigma^2$, i.e., $\boldsymbol{\epsilon} \sim \mathcal{CN}(\mathbf{0}, \sigma^2\mathbf{I})$. After transmitting over a noisy channel, the noisy symbols received by the receiver can be expressed by
 \begin{align}
     \hz = \z + \boldsymbol{\epsilon}.
 \end{align}
  The considered AWGN channel can be fully characterized by the channel SNR $\omega \in \Omega$, where $\Omega$ is a set of possible values of $\omega$\footnote{Although we assume an AWGN channel and use SNR {to represent the} channel conditions for simplicity, this model can be extended to other channel models as long as the channel conditions can be estimated by the receiver and fedback to the transmitter.}.  
The condition $\omega$ is {assumed to be} perfectly estimated at the receiver side and {fedback} to the transmitter. Formally, the channel effect can be represented by the conditional distribution $p_{\omega}(\hz|\z)$.
At the receiver side, the decoder $p_{\bTheta}(\hat{\x}|\hz)$ leverages the received noisy signal $\hz$ to produce a reconstruction of the input $\hat{\x} \in \mathcal{X}$, where $\bTheta$ is the trainable and adjustable parameters of the decoder. After introducing the above random variables, we can establish the following Markov chain to represent the whole encoding and decoding process,
 \begin{align}
		X \stackrel{\bPhi}{\longleftrightarrow} Z \stackrel{\omega}{\longleftrightarrow} \hZ \stackrel{\bTheta}{\longleftrightarrow} \hat{X} .\label{eq:markov_chain}
	\end{align}
 Note that this system model can be generalized to task-oriented communication by introducing the target variable $Y$ into the data source $p(\x, \y)$. In particular, the decoder $p_{\bTheta}(\hat{\y}|\hz)$ just leverages $
 \hz$ and outputs the target estimation $\hy \in \mathcal{Y}$. Then, we can also establish the following Markov chain for the encoding and decoding process,
 \begin{align}
		Y&\longleftrightarrow X \stackrel{\bPhi}{\longleftrightarrow} Z \stackrel{\omega}{\longleftrightarrow} \hZ \stackrel{\bTheta}{\longleftrightarrow} \hat{Y} .\label{eq:markov_chain_task}
	\end{align}
 \subsection{Problem Formulation}
 With the system model illustrated in Fig.~\ref{fig:system-model_Hyper}, our objective is to develop a scalable and computationally efficient framework for learning the optimal DJSCC schemes for various channel conditions in a single training round. Specifically, given the encoder $p_{\bPhi}(\z|\x)$, the channel model $p_{\omega}(\hz|\z)$, and the decoder $p_{\bTheta}(\hat{\mathbf{x}}|\hz)$, the objective of DJSCC is to minimize the following objective function, 
 \begin{align}
     \mathcal{L}(\bPhi, \bTheta; \omega) &= \mathbb{E}_{p(\x)}\left [\mathbb{E}_{p_{\bPhi}(\z|\x)p_{\omega}(\hz|\z)}\left [-\log p_{\bTheta}(\x|\hz)\right ]\right ]. \label{loss: original}
 \end{align}
 The distortion $\mathbb{E}_{p_{\bPhi}(\z|\x)p_{\omega}(\hz|\z)}[-\log p_{\bTheta}(\x|\hz)]$ in $\eqref{loss: original}$ is expressed in a general probabilistic format. The decoder $p_{\bTheta}(\x|\z)$ is based on a spherical Gaussian observation model, and the distortion is measured as the mean squared error (MSE) between the input data $\x$ and the reconstruction $\hat{\mathbf{x}}$. For task-oriented communication schemes with probability model~\eqref{eq:markov_chain_task}, the objective function $\mathcal{L}(\bPhi, \bTheta; \omega)$ becomes,
  \begin{align}
     \mathcal{L}(\bPhi, \bTheta; \omega) &= \mathbb{E}_{p(\x, \y)}\left [\mathbb{E}_{p_{\bPhi}(\z|\x)p_{\omega}(\hz|\z)}\left[-\log p_{\bTheta}(\y|\hz)\right]\right ]. \label{loss: original_task}
 \end{align}
If the target $\y \in \{ 1, 2, \dots, K\}$ is classified into $K$ classes and the decoder $p_{\bTheta}(\y|\hz)$ outputs {the} probability vectors for the $K$ possible classes, the distortion is calculated as the cross-entropy.  

 Given the channel condition $\omega$, the adaptive DJSCC schemes adaptively adjust their parameters to the optimal parameters $\bPhi^*(\omega)$ and $\bTheta^*(\omega)$ that minimize the objective $\mathcal{L}(\bPhi, \bTheta; \omega)$. {That is,}
 \begin{align}
     \bPhi^*(\omega), \bTheta^*(\omega) = \argmin\limits_{\bPhi, \bTheta} \mathcal{L}(\bPhi, \bTheta; \omega).
 \end{align} 
In the next section, we will propose a single model for DJSCC schemes that adapts to any channel conditions $\omega$ within one training round. 

\section{Adaptive DJSCC using Hypernetworks}\label{sec: proposed_method}
In this section, we first develop a general framework to achieve adaptive DJSCC by utilizing hypernetworks. These hypernetworks generate optimal parameters $(\bPhi^*(\omega), \bTheta^*(\omega))$ for the given channel conditions $\omega$. Subsequently, we propose a scalable and memory-efficient hypernetwork parameterization to reduce the additional storage space, named Hyper-AJSCC.
\subsection{Hypernetworks for DJSCC}

The concept of hypernetwork was proposed in {the} seminal work~\cite{ha2016hypernetworks} to utilize a single neural network to generate the parameters for other networks called target networks. 
This approach offers several benefits, including weight sharing, dynamic architecture, and adaptive neural networks. By using hypernetworks, we can avoid the need for computationally expensive training processes of multiple neural networks for different scenarios.
This key characteristic of hypernetworks provides us with the opportunity to develop adaptive DJSCC schemes tailored to various channel conditions, without necessitating the retraining or storage of multiple neural networks.

\begin{figure}[t]
		\centering
		\includegraphics[width=0.93\linewidth]{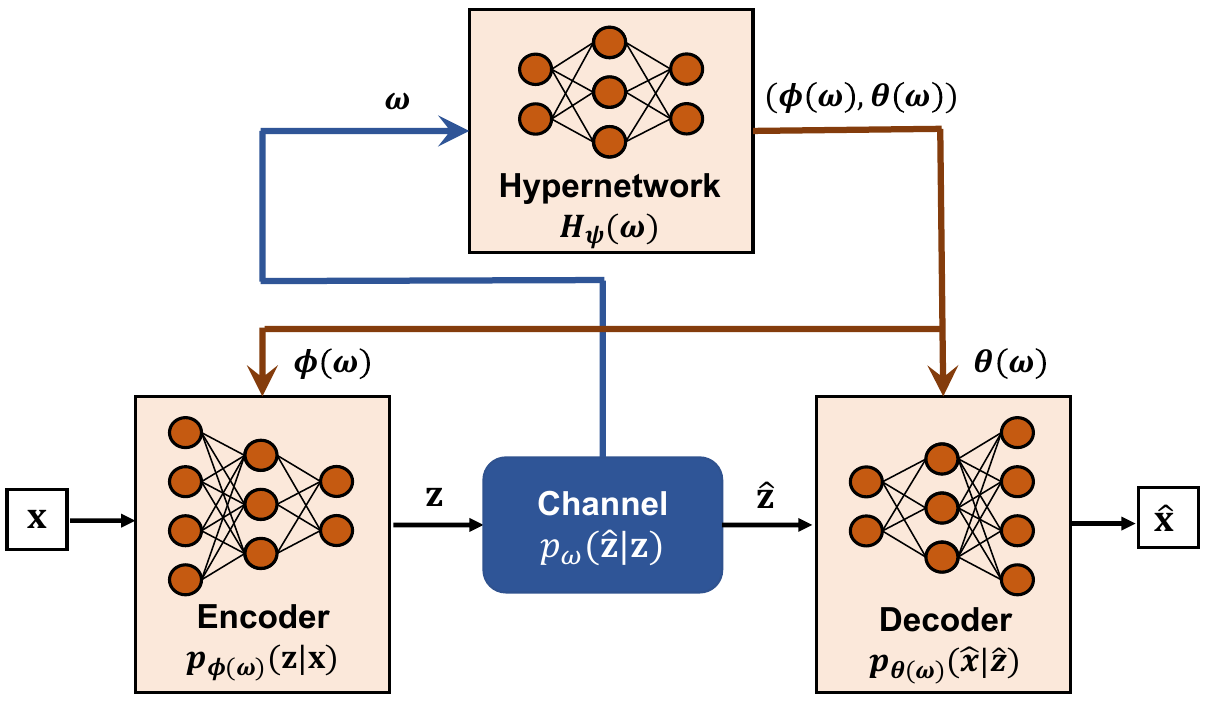}
		\caption{The proposed hypernetwork-based framework: A hypernetwork $H$ takes channel conditions $\omega$ as input and generates the parameters $\bPhi(\omega)$ and $\bTheta(\omega)$ of the encoder and decoder.
		}
		\label{fig:general-hyper}
	\end{figure}

The proposed framework based on hypernetworks is illustrated in Fig.~\ref{fig:general-hyper}. We regard the encoder and decoder as the target networks and propose to utilize a hypernetwork $H_{\bPsi}$ to generate their parameters $\bPhi(\omega)$ and $\bTheta(\omega)$ with the input of {the} channel condition $\omega$ as
\begin{align}
 (\bPhi(\omega), \bTheta(\omega)) &= H_{\bPsi}(\omega),
\end{align}
where $H_{\bPsi}: \Omega \to \Phi \times \Theta$ that maps the channel condition $\omega$ to $(\bPhi(\omega), \bTheta(\omega))$, where $\bPsi$ represents the meta parameters of the hypernetwork.

In the conventional DJSCC methods, the parameters $(\bPhi, \bTheta)$ of the encoder and decoder are directly optimized in an end-to-end manner. However, in the proposed hypernetwork-based method, we instead optimize the hypernetwork $H_{\psi}$ to generate optimal solutions $(\bPhi^*(\omega), \bTheta^*(\omega))$ that minimize the objective function $\mathcal{L}(\bPhi(\omega), \bTheta(\omega); \omega)$.
In particular, the meta parameters $\bPsi$ of {the} hypernetwork $H_{\bPsi}$ is trained to minimize the following objective function 
\begin{align}
    \mathcal{H}(\bPsi) = \mathbb{E}_{p(\omega)}[\mathcal{L}(H_{\bPsi}(\omega);\omega)],
\end{align}
where the channel condition $\omega$ is sampled from a prior distribution $p(\omega)$ over $\Omega$ during training. 
Then, the optimal parameters $\bPhi^*(\omega)$ and $ \bTheta^*(\omega)$ are obtained by the optimized hypernetwork:
\begin{align}
    \bPhi^*(\omega), \bTheta^*(\omega) = H_{\bPsi^*}(\omega), \label{eq:hyper_loss}
\end{align}
where $\bPsi^* = \argmin\limits_{\bPsi} \mathcal{H}(\bPsi)$ is the optimized meta parameters of the hypernetwork.
\subsection{Hyper-AJSCC}
Directly applying hypernetworks into the DJSCC requires additional storage space for the deployment of the introduced hypernetwork $H_{\bPsi}$ and the additional step to generate the parameters $\bPhi(\omega)$ and $\Theta(\omega)$ for the encoder and decoder. Inspired by the applications of hypernetworks in hyperparameter optimization~\cite{bae2022multi}, we integrate the hypernetwork into the encoder and decoder by designing a memory-efficient and scalable hypernetwork parameterization. The proposed adaptive DJSCC scheme is named Hyper-AJSCC, which can be easily extended to other DJSCC networks by introducing very limited additional parameters and without destroying the consistency of the original backbone of neural networks. 
 
Firstly, we consider the $l$-th layer of the DJSCC network, which takes the feature $\mathbf{f}^{(l)} \in \mathbb{R}^{D_{l}}$ produced from the previous layer and outputs the feature $\mathbf{f}^{(l+1)} \in \mathbb{R}^{D_{l+1}}$. As the trainable parameters of {the} $l$-th layer, weight and bias, can be expressed as $\W^{(l)}\in \bR^{D_{l+1} \times D_l}$ and $\bb^{(l)} \in \bR^{D_{l+1}}$, respectively. Thus, the obtained features from {the} $l$-th layer are expressed by,
\begin{align}
    \mathbf{f}^{(l+1)} &= \sigma^{(l)}(\W^{(l)}\mathbf{f}^{(l)} + \bb^{(l)}) \label{eq: original_activation},
\end{align}
where $\sigma^{(l)}(\cdot)$ denotes the activation function of the $l$-th layer. 

\begin{figure}[t]
		\centering
		\includegraphics[width=\linewidth]{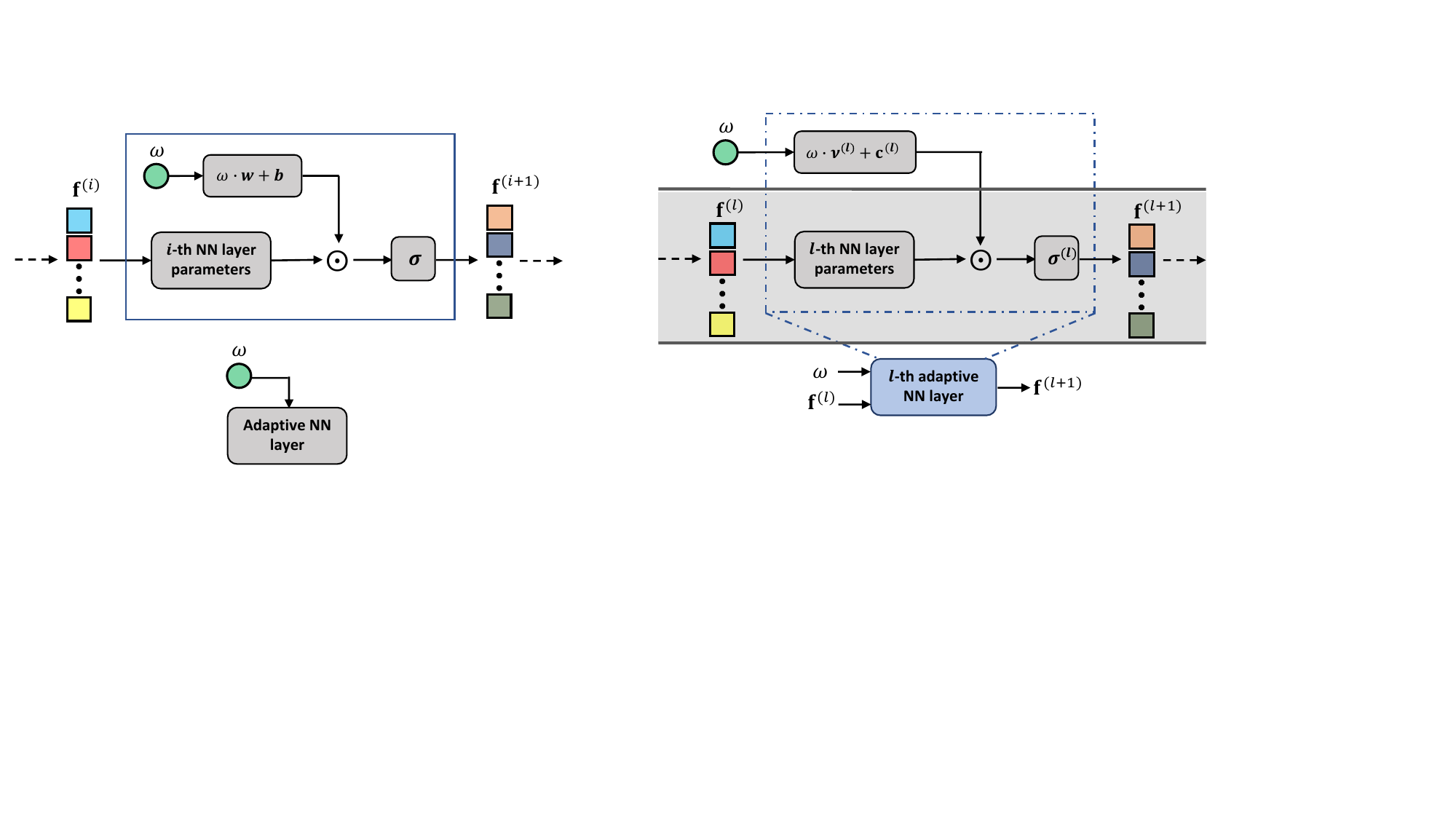}
		\caption{The proposed hypernetwork parameterization for {the} $l$-th layer of DJSCC schemes. The gray panel represents the basic module, which maintains the original neural network backbone.
		}
		\label{fig:method-Hyper}
	\end{figure}

Next, we present how to model the adaptive parameters of {the} $l$-th layer, $\W^{(l)}(\omega)$ and $\bb^{(l)}(\omega)$, by using the following hypernetwork parameterization
\begin{align}
    \W^{(l)}(\omega) & = (\omega \cdot \bNu^{(l)}+ \bc^{(l)} ) \odot_{\text{row}} \W^{(l)}_0,\\
    \bb^{(l)}(\omega) & = (\omega \cdot \bNu^{(l)}+ \bc^{(l)} ) \odot \bb^{(l)}_0,
\end{align}
where $\odot$ and $\odot_{\text{row}}$ denote the element-wise multiplication for vectors and row-wise multiplication for matrices, respectively. Note that the meta parameters for {the} $l$-th layer are $\bPsi^{(l)} = \{ \bNu^{(l)}, \mathbf{c}^{(l)}, \W_0^{(l)}, \bb^{(l)}_0 \}$. To develop a memory-efficient and scalable hypernetwork-based scheme, we further derive the features $\mathbf{f}^{(l+1)}$ produced by the adaptive $l$-th layer as follows,
\begin{align}
    \f^{(l+1)} (\omega) &= \sigma^{(l)}(\W^{(l)}(\omega) \f^{(l)} + \bb^{(l)}(\omega))\\
    &= \sigma^{(l)}(\underbrace{(\omega \cdot \bNu^{(l)}+ \bc^{(l)} )}_{\text{Element-wise scaling}} \odot \underbrace{(\W^{(l)}_0\mathbf{f}^{(l)} + \bb^{(l)}_0)}_{\text{Basic module}}) \label{eq: activation}.
\end{align}
As shown in \eqref{eq: activation}, we observe that the proposed hypernetwork-based network architecture can be decomposed into two main parts, \emph{Element-wise scaling} and \emph{Basic module}. Specifically, the basic module has the same form as the original neural network backbone in \eqref{eq: original_activation}, and the element-wise scaling can be easily achieved with a linear transformation. Therefore, the hypernetwork for {the} $l$-th adaptive layer can be directly implemented by adding a linear transformation on the existing module, as illustrated in Fig.~\ref{fig:method-Hyper}.
 
The above method is based on fully connected architecture, but the proposed hypernetwork parameterization can be easily extended to other neural network architectures such as convolutional layers and resnet blocks as follows.
\paragraph{Convolutional layers} We consider {the} $l$-th convolutional layer, which receives input features $\f^{(l-1)} \in \bR^{c_{l-1}\times h_{l-1}\times w_{l-1}}$ and is equipped with filters characterized by kernels $\mathbf{C}^{(l)} \in \bR^{c_{l-1}\times K_i\times K_i}$ and bias $\mathbf{b}^{(l)} \in \bR$. Analogous to the aforementioned parameterization, the adaptive parameters $\mathbf{C}^{(l)}(\omega)$ and $\mathbf{b}^{(l)}(\omega)$ are formulated as follows:
\begin{align}
    \mathbf{C}^{(l)}(\omega) & = (\omega \cdot \bNu^{(l)}+ \bc^{(l)} ) \odot_{\text{row}} \mathbf{C}^{(l)}_0,\\
    \bb^{(l)}(\omega) & = (\omega \cdot \bNu^{(l)}+ \bc^{(l)} ) \odot \bb^{(l)}_0.
\end{align}
\paragraph{Resnet blocks} The parameterization for ResNet blocks can be readily extended by applying the above parameterization techniques to all convolutional layers within the ResNet blocks.

\subsection{Training Algorithm}
\addtolength{\topmargin}{0.05in}
By employing the proposed hypernetwork parameterization, the Hyper-AJSCC scheme can effectively incorporate and adapt to different channel conditions. 
Next, we aim to develop a training process for Hyper-AJSCC, which is expected to learn a mapping from any channel conditions $\omega$ to the optimal parameters $(\bPhi^*(\omega), \bTheta^*(\omega))$. Thus, we take a prior distribution $p(\omega)$ over $\Omega$ and train Hyper-AJSCC to minimize the loss $\mathcal{H}(\bPsi)$ presented in \eqref{eq:hyper_loss} by sampling $\omega \sim p(\omega)$. Specifically, by sampling $L$ realizations of $\omega$ from $p(\omega)$, we can form the Monte Carlo estimation $\Tilde{\mathcal{H}}(\bPsi) \simeq \mathcal{H}(\bPsi)$
\begin{align}
    \Tilde{\mathcal{H}}(\bPsi) &= \frac{1}{L}\sum\limits_{l=1}^L\mathcal{L}(\bPhi^{(l)}, \bTheta^{(l)}; \omega^{(l)}),
\end{align}
where $(\bPhi^{(l)}, \bTheta^{(l)}) = H_{\bPsi}(\omega^{(l)})$ and $ \omega^{(l)} \sim p(\omega)$. 
The training process for Hyper-AJSCC is outlined in Algorithm~\ref{algo-Hyper-AJSCC}. 
Although adopting a prior distribution $p(\omega)$ that aligns with real-world channel condition distributions can enhance the training process, it is challenging to obtain such distribution. 
In our empirical evaluation, we demonstrate that Hyper-AJSCC can achieve optimal performance by simply utilizing a uniform prior $p(\omega)$.
\begin{algorithm}[t]
\small
\caption{Training Hyper-AJSCC}
\begin{algorithmic}[1]
\label{algo-Hyper-AJSCC}
\REQUIRE $T$ (number of epochs), batch size $N$, sampling distribution for channel condition $p(\omega)$.
\WHILE{epoch $t=1$ to $T$}
    \STATE Sample a mini-batch of data samples $\{\x^{(i)}\}_{i=1}^{N}$
    \STATE Sample a mini-batch of channel conditions $\{ \omega^{(i)} \}_{i=1}^N \sim p(\omega)$
    \STATE Generate channel models $\{p_{\omega^{(i)}}(\hz|\z)\}_{i=1}^N$ according to $\{ \omega^{(i)} \}_{i=1}^N$
    \WHILE{$i=1$ to $N$}
        \STATE Compute $\z^{(i)}$ by inputting $\x^{(i)}$ and $\omega^{(i)}$ to the encoder
        \STATE Estimate the received symbols $\hz^{(i)}$ from $p_{\omega^{(i)}}(\hz|\z)$
        \STATE Estimate the outputs by inputting $\hz^{(i)}$ and $\omega^{(i)}$ to the decoder
    \ENDWHILE
    \STATE Compute the loss $ \Tilde{\mathcal{H}}(\bPsi)$ and update the parameters $\bPsi$ through backpropagation.
\ENDWHILE
\end{algorithmic}
\end{algorithm}
\section{Experiments}
In this section, we evaluate the performance of the proposed hyper-DJSCC in data/task-oriented communications by performing a series of experiments on image transmission and image classification tasks\footnote{The source code of the proposed Hyper-AJSCC is available at: https://github.com/SongjieXie/Hyper-AJSCC.}.
\subsection{Implementation Details}
\subsubsection{Baselines and Dataset}
\begin{table}
		\caption{The BDJSCC Network Architecture for image transmission}
		\label{tab:image}
		\begin{tabular}{p{0.2\columnwidth}|p{0.45\columnwidth}|p{0.2\columnwidth}}
			\toprule
			& \textbf{Layer} & \textbf{Output} \\
			\midrule
			\multirow{2}{*}{\textbf{Decoder}} & Conv $\times$ $4$ & $256 \times 8 \times 8 $\\
                                                  & Conv $+$ Power normalization & $d$ \\
			\midrule
			\multirow{2}{*}{\textbf{Decoder}} & \text{Deconv} $\times$ $4$  & $256 \times 16 \times 16$\\
			& \text{Deconv} $+$ \text{Tanh}& $3 \times 32 \times 32$\\
			\bottomrule
		\end{tabular}
	\end{table}
	\begin{table}[t]
		\caption{The BDJSCC Network Architecture for Image Classification}
		\label{tab:classification}
		\begin{tabular}{p{0.13\columnwidth}|p{0.49\columnwidth}|p{0.22\columnwidth}}
			\toprule
			& \textbf{Layer} & \textbf{Outputs} \\
			\midrule
			\multirow{4}{*}{\textbf{Encoder}} & Conv $\times$ $2$ $+$ ResBlock & $256\times 16\times 16$\\
			& Conv $\times$ $3$ & $4 \times 4 \times 4$\\
                & Reshape $+$ Fully-connected layer & $d$\\
			\midrule
			\multirow{4}{*}{\textbf{Decoder}}& Fully-connected Layer $\times$ $3$  & $4\times 4 \times 4$\\
			& Conv $+$ Resblock                   & $512 \times 4 \times 4$\\
                & Pooling layer & $512$\\
			& Fully-connected $+$ Softmax & $10$\\
			\bottomrule
		\end{tabular}
	\end{table}
We select representative DJSCC schemes for image transmission and cooperative inference as the basic DJSCC (BDJSCC) neural network architectures. The corresponding BDJSCC network architectures for image transmission~\cite{kurka2020deepjscc} and cooperative inference~\cite{shao2021learning} are shown in Table~\ref{tab:image} and Table~\ref{tab:classification}, respectively. Moreover, we adopt the neural network architecture of BDJSCC as the backbone of our proposed Hyper-AJSCC for {a} fair comparison.
All experiments are conducted using the benchmark dataset \emph{CIFAR-10}~\cite{krizhevsky2009learning}, which consists of a collection of $32\times 32$ color images categorized into 10 classes. The dataset includes a training set with 50,000 images and a test set with 10,000 images.

\subsubsection{Performance Metrics}
Following the DJSCC literature~\cite{bourtsoulatze2019deep, xu2021wireless}, we use the compression ratio $R= d/n$ to characterize the communication efficiency of the system, where $n$ and $d$ are the source bandwidth (the dimension of $\x$) and channel bandwidth (the dimension of $\z$), respectively. The performance is evaluated by the peak signal-to-noise ratio (PSNR), which is a standard performance metric for image reconstruction. The PSNR is defined as, 
\begin{align}
\text{PSNR} = 10\log_{10}\frac{\text{MAX}(\mathbf{x})^2}{\text{MSE}(\mathbf{x}, \hat{\mathbf{x}})},
\end{align}
where $\text{MAX}(\mathbf{x})$ represents the maximum pixel value in the image $\mathbf{x}$, and $\text{MSE}(\mathbf{x}, \hat{\mathbf{x}})$ denotes the MSE between the original image $\mathbf{x}$ and the reconstructed image $\hat{\mathbf{x}}$. 
The experiments on task-oriented communications are performed with classification tasks using the CIFAR-10 benchmark datasets. For a fair comparison, the channel symbols $\z$ encoded by all the evaluated methods have the
same dimensionality $d$, and the standard metric of top-1 accuracy is employed to evaluate the classification results. 

\subsection{Experimental Results on Data-Oriented Communications}
 \begin{figure}
		\centering
		
		\subfloat[]{
			\centering
			\includegraphics[width=0.78\linewidth]{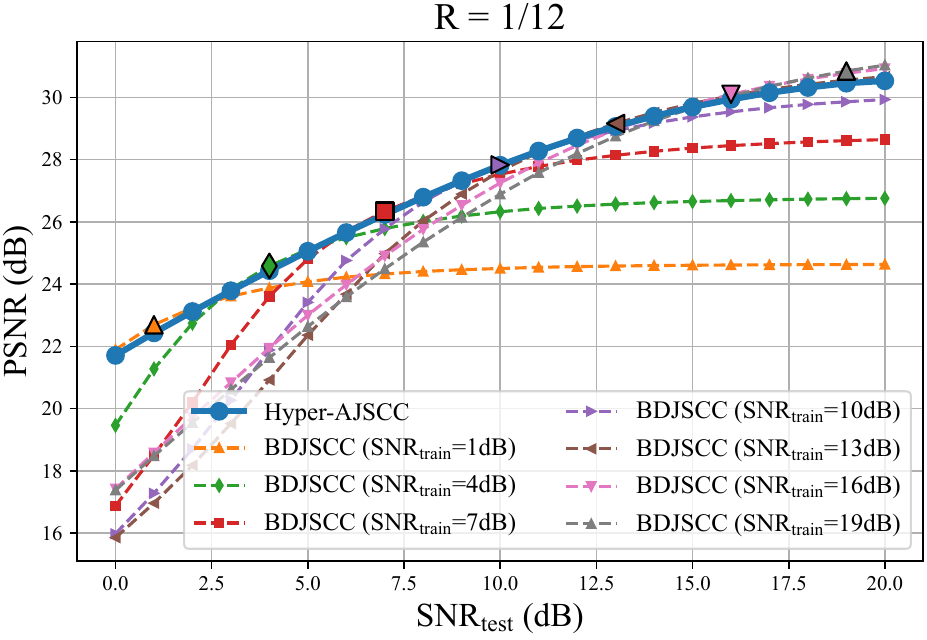}
		}
  
		\subfloat[]{
			\centering
			\includegraphics[width=0.78\linewidth]{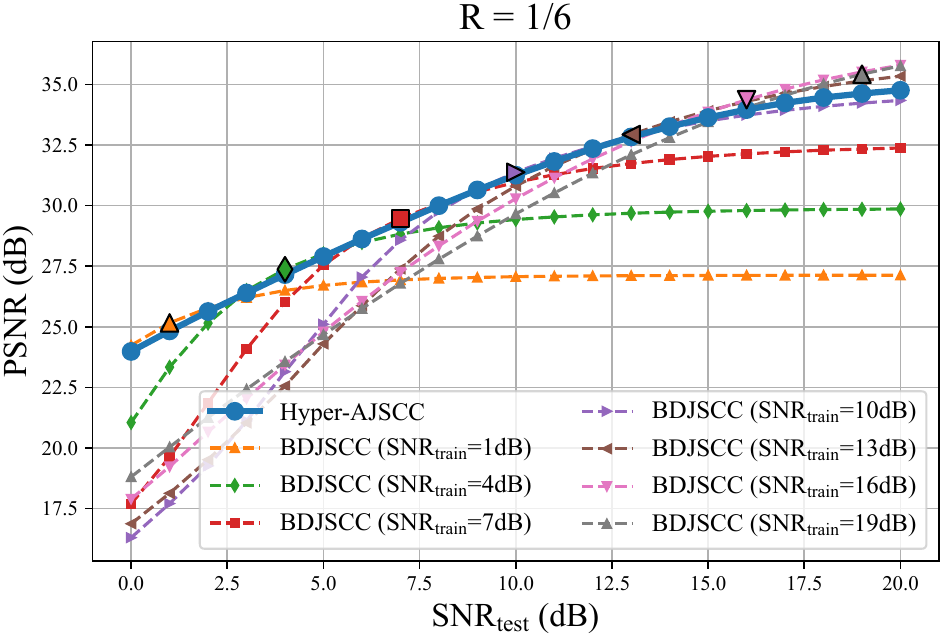}
		}
		\caption{Performance of the proposed Hyper-AJSCC compared to baseline BDJSCCs under varying training SNR with compression ratios (a) $R=1/12$ and (b) $R=1/6$. The outlined markers represent the performance of BDJSCCs when the test SNR matches their training SNR, i.e., $\text{SNR}_{\text{train}} = \text{SNR}_{\text{test}}$.
		}
		\label{fig:recon}
	\end{figure}
 We first evaluate the effectiveness of our proposed method on image transmission.
 All BDJSCC schemes are trained at specific SNRs, denoted as $\textrm{SNR}_{\textrm{train}} \in \{1\textrm{dB}, 4\textrm{dB}, 7\textrm{dB}, 10\textrm{dB}, 13\textrm{dB}, 16\textrm{dB}, 19\textrm{dB} \}$. {In contrast}, our Hyper-AJSCC scheme is trained under the uniform distribution of $\textrm{SNR}_\textrm{train} \in [0\textrm{dB}, 20\textrm{dB}]$. Subsequently, we evaluate the performance of BDJSCCs and Hyper-AJSCC over a range of SNRs varying from $0\textrm{dB}$ to $20\textrm{dB}$. The experimental results shown in Fig.~\ref{fig:recon} indicate the performance with compression ratios $R=1/12$ and $R=1/6$, respectively. 
 The outlined markers represent the optimal performance of BDJSCCs when the $\textrm{SNR}_\textrm{test}$ is the same for the testing SNR (i.e., $\textrm{SNR}_\textrm{train} = \textrm{SNR}_\textrm{test}$). We observe that Hyper-AJSCC achieves the optimal performance of BDJSCCs trained with different values of $\textrm{SNR}_\textrm{train}$, except for a small performance gap in the high SNR regime. This indicates that our method adapts well to SNR variations across different SNR values.

\begin{figure}[t]
		\centering
		\includegraphics[width=0.7\linewidth]{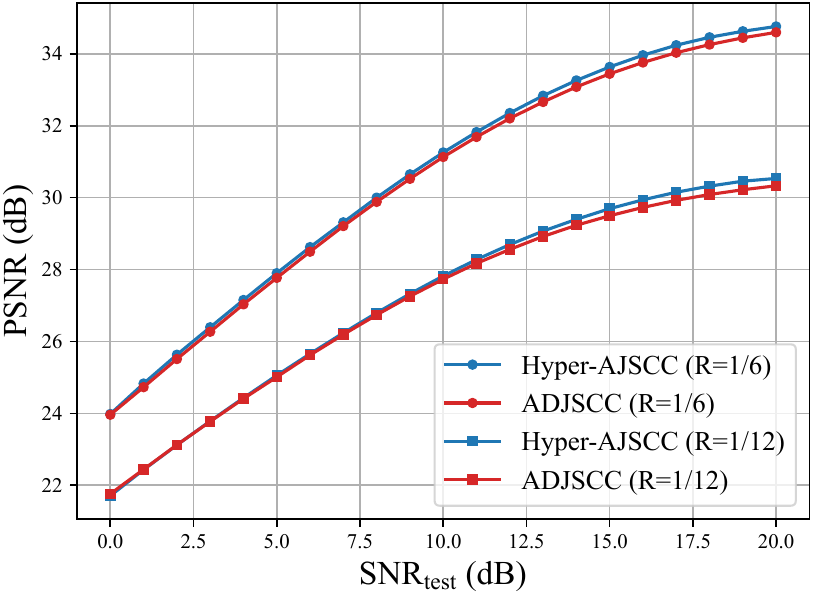}
		\caption{Performance of the proposed Hyper-AJSCC compared to ADJSCC with different compression ratios $R$.
		}
		\label{fig:recon-compare}
	\end{figure}

We further compare Hyper-AJSCC with the state-of-the-art method, ADJSCC~\cite{xu2021wireless}, under {an} identical setting\footnote{To ensure a fair comparison in the experiments of image transmission, the ADJSCC, as it originally proposed in \cite{xu2021wireless}, adopts {an} identical neural network architecture as BDJSCC. Consequently, BDJSCC, ADJSCC, and our Hyper-AJSCC share the same neural network backbone. }.
As illustrated in Fig.~\ref{fig:recon-compare}, Hyper-AJSCC shows a slight performance improvement when compared to ADJSCC, especially in the high-SNR regime. 
Furthermore, we compute the additional storage overhead introduced by the ADJSCC and Hyper-AJSCC models on the basis of the BDJSCC backbone in Table.~\ref{tab:storage}. 
It is shown that our Hyper-AJSCC scheme only introduces less than one-tenth of the parameters introduced by ADJSCC. 
Because of the small additional storage overhead of Hyper-AJSCC, it is easy to extend our framework to various existing DJSCC schemes to enable their adaptation to channel variations.
 	\begin{table}
  \centering
		\caption{Introduced Parameters and Additional Storage Requirements for Evaluated Methods}
		\label{tab:storage}
			\begin{tabular}{ccc}
				\toprule
				\textbf{Method} & \textbf{\# parameters} & \textbf{Storage} \\
				\midrule
				ADJSCC &67840& 265 \text{KB}\\
				Hyper-AJSCC&\textbf{4118}& \textbf{16 \text{KB}}\\
				\bottomrule
		\end{tabular}
	\end{table}
 \subsection{Experimental Results on Task-Oriented Communications}
 \begin{figure*}
		\centering
		
		\subfloat[]{
			\centering
			\includegraphics[width=0.41\linewidth]{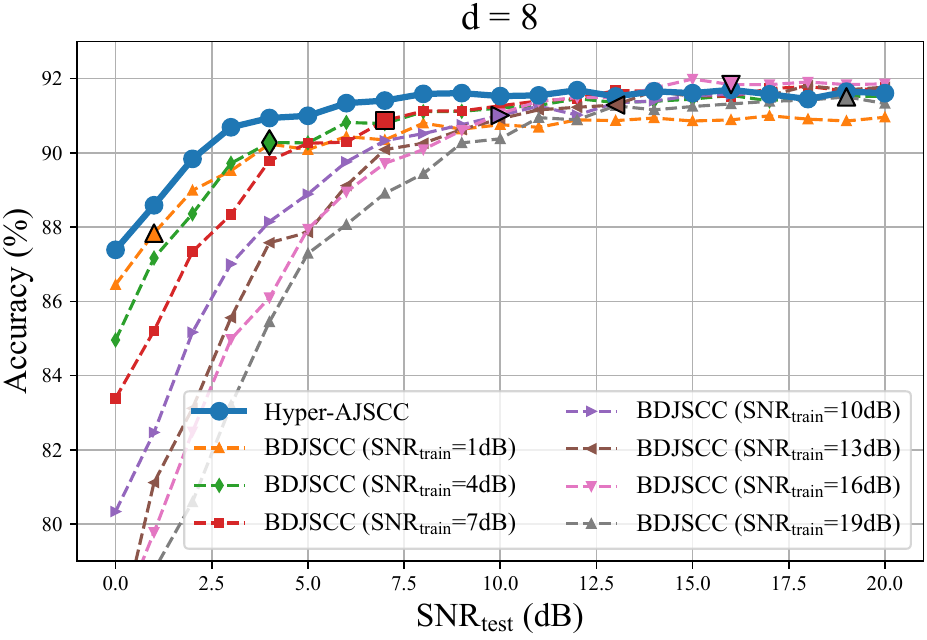}
		}
		\subfloat[]{
			\centering
			\includegraphics[width=0.40\linewidth]{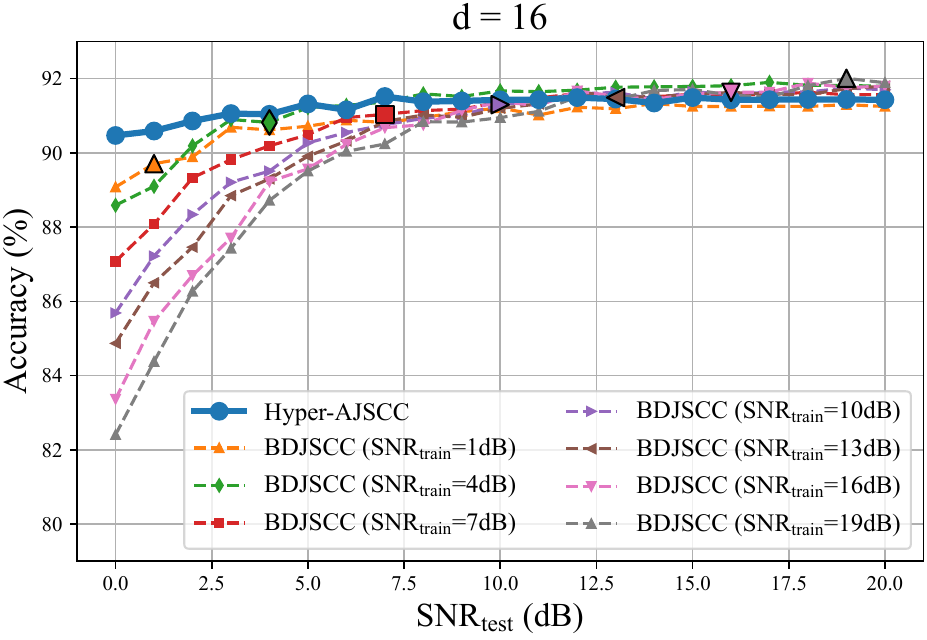}
		}
		\caption{Performance of the proposed Hyper-AJSCC compared to baseline BDJSCCs under varying training SNR for image classification tasks, with (a) $d=8$ and (b) $d=16$.
		}
		\label{fig:task}
	\end{figure*}
To further verify the performance of our method in task-oriented communications, we apply the proposed Hyper-AJSCC scheme to cooperative inference. 
Following the same experimental setup employed for data-oriented communications experiments, we show the experimental results in Fig.~\ref{fig:task}. We observe that the proposed Hyper-AJSCC scheme outperforms the BDJSCC schemes, particularly when the value of $\textrm{SNR}_\textrm{test}$ is relatively small. 
Our proposed Hyper-AJSCC can outperform the performance of BDJSCC with perfect SNR  (e.g., the makers) except in the high SNR regime.
This is because channel adaptability in task-oriented communication is more intricate due to the inherent tradeoff between informativeness and robustness~\cite{10159007}, and there does not exist an overall optimal performance for channel-adaptive DJSCC schemes in task-oriented communications.

\section{Conclusions}
In this work, we investigated the channel adaptability problems for DJSCC and proposed a generic framework to achieve channel-adaptive DJSCC by leveraging hypernetworks. Subsequently, we proposed {a} hypernetwork-based adaptive DJSCC and further proposed a scalable and memory-efficient scheme by introducing a limited number of hyperparameters into each layer of the neural network. The proposed framework can be seamlessly integrated into any conventional network architecture of DJSCC. 
Extensive experiments demonstrated the superior performance and less storage requirement of the Hyper-AJSCC compared to state-of-the-art baselines on the task of image transmission and cooperative inference. Our work {has demonstrated} the potential of hypernetworks in adaptive DJSCC, and a promising future direction is to further explore the hypernetwork-based framework for incorporating various data distributions, thus enabling the development of DJSCC schemes that are adaptive to both data sources and channels.


\linespread{0.96}{
\bibliography{IEEEabrv, ref}
\bibliographystyle{IEEEtran}
}
\end{document}